hep-ph/9508387  8 Nov 1995

# On the Measurability of the Structure Function $g_1(x, Q^2)$ in $ep$ collisions at HERA

Johannes Blümlein

*DESY – Zeuthen,
Platanenallee 6, D–15735 Zeuthen, Germany*

**Abstract**

The possibility is investigated to measure the polarized structure function $g_1(x, Q^2)$ in the collider mode of HERA operating with a polarized lepton and proton beam. The $x$ dependence of $g_1$ can be measured at a statistical precision of $\sim 20\%$ to $80\%$ in the range $0.0005 < x < 0.5$ correlated to virtualities $15 < \langle Q^2 \rangle < 3500\,\mathrm{GeV}^2$ at beam polarizations $\lambda_p \sim \lambda_e = 0.8$ and $\mathcal{L}_{int} = 60\mathrm{pb}^{-1}$. At a low energy option, $E_p = 300\,\mathrm{GeV}$, the statistical accuracy improves to values of $\sim 15\%$ to 50 % in the range $10^{-3} < x < 0.5$ and $10 < Q^2 < 800\,\mathrm{GeV}^2$.

# 1 Introduction

The possibility to study deep inelastic electron–proton scattering with both polarized electron and proton beams at the HERA collider would open up a new era in the investigation of the nucleon spin structure [1]. Far smaller values of Bjorken $x$ than ever probed in fixed target experiments would become accessible and the behaviour of polarized structure functions in the range down to $x \sim 0.0001$ could be investigated. This is of particular importance since the present analysis of sum–rules relies on extrapolations in this range only. Moreover little is known on the scaling violations of polarized structure functions from experiment in the whole $x$ range so far. Due to the kinematical domain of HERA a far wider $Q^2$ range can be accessed. Furthermore both charged and neutral current reactions can be studied from which constraints on the flavour structure of polarized structure functions may be obtained.

In the present paper we concentrate on the case of neutral current deep inelastic scattering in the $Q^2$ range dominated by photon exchange, i.e. $Q^2 \lesssim 1000$ GeV$^2$ [2]. Here the two polarized structure functions $g_1(x,Q^2)$ and $g_2(x,Q^2)$ and the two unpolarized structure functions $2xF_1(x,Q^2)$ and $F_2(x,Q^2)$ determine the polarization asymmetries. We analyse the possibilities to measure the polarization asymmetries $A_\parallel$ and $A_\perp$ in the HERA collider mode. An estimate is given of the statistical accuracy to which the structure function $g_1(x,Q^2)$ can be measured.

# 2 Polarization Asymmetries

The structure functions $g_1(x,Q^2)$ and $g_2(x,Q^2)$ can be determined from the measurement of the polarization asymmetries $A_\parallel$ and $A_\perp$ ([3, 4]):

$$A_\parallel = \frac{d^2\sigma^{\Rightarrow} - d^2\sigma^{\Leftarrow}}{d^2\sigma^{\Rightarrow} + d^2\sigma^{\Leftarrow}} \tag{1}$$

$$A_\perp = \frac{d^2\sigma^{\Uparrow} - d^2\sigma^{\Downarrow}}{d^2\sigma^{\Uparrow} + d^2\sigma^{\Downarrow}} \tag{2}$$

with

$$\frac{d^2\sigma^{\Rightarrow}}{dxdy} - \frac{d^2\sigma^{\Leftarrow}}{dxdy} = -\lambda_e \lambda_p \frac{8\pi\alpha^2}{Q^2} \left\{ \left[ 2 - y - \frac{2M^2 xy}{S} \right] g_1(x,Q^2) - \frac{4M^2 x}{S} g_2(x,Q^2) \right\} \tag{3}$$

$$\frac{d^3\sigma^{\Uparrow}}{dxdyd\phi} - \frac{d^3\sigma^{\Downarrow}}{dxdyd\phi} = \lambda_e \lambda_p \frac{4\alpha^2}{Q^2} \cos\phi \left[ \frac{4M^2 x(1-y)}{yS} \right]^{1/2} \left[ 1 - \frac{M^2 xy}{(1-y)S} \right]^{1/2}$$
$$\times \left\{ y g_1(x,Q^2) + 2 g_2(x,Q^2) \right\} \tag{4}$$

and

$$\frac{d^2\sigma^{\Rightarrow}}{dxdy} + \frac{d^2\sigma^{\Leftarrow}}{dxdy} \equiv \frac{d^3\sigma^{\Uparrow}}{dxdy} + \frac{d^3\sigma^{\Downarrow}}{dxdy}$$
$$= \frac{4\pi\alpha^2 S}{Q^4} \left\{ y^2 2xF_1(x,Q^2) + 2\left(1 - y - \frac{xyM^2}{S}\right) F_2(x,Q^2) \right\}. \tag{5}$$

Here $\rightarrow$ denotes the orientation of the electron and proton ($\Rightarrow$) polarization. $x$ and $y$ are the Bjorken variables, $S = 4E_e E_p$, $Q^2 = xyS$, $M$ is the proton mass, and $\lambda_e$ and $\lambda_p$ are the electron and proton beam polarizations. $\phi$ denotes the angel between the planes ($\boldsymbol{k}$, $\boldsymbol{k}'$) and ($\boldsymbol{k}$, $\boldsymbol{s}$), where $\boldsymbol{k}$, $\boldsymbol{k}'$, and $\boldsymbol{s}$ are the 3-momenta of the incoming and outgoing electron, and the proton spin, respectively.

The contribution due to the structure function $g_2(x,Q^2)$ in $A_\parallel$ is suppressed by a factor of $\sim 4M^2x/(S(2-y))$ relative to $g_1(x,Q^2)$. In the HERA collider mode this term is very small and can be disregarded as well as other terms of $\mathcal{O}(M^2/S)$ leading to

$$A_\parallel = -\lambda_e \lambda_p \frac{Y_- 2xg_1(x,Q^2)}{Y_+ 2xF_1(x,Q^2) - 2(1-y)F_L(x,Q^2)}, \qquad (6)$$

where $F_L(x,Q^2) = F_2(x,Q^2) - 2xF_1(x,Q^2)$, and $Y_\pm = 1 \pm (1-y)^2$.

The asymmetry $A_\perp$ contains a factor $f_{A_\perp} = 4M/\sqrt{S}$ leading to a strong suppression in the case of the collider mode at HERA in comparison with the kinematics in fixed target experiments at SLAC or HERMES.

$$A_\perp = \lambda_e \lambda_p \frac{\cos\phi}{2\pi} f_{A_\perp} \sqrt{xy(1-y)} \frac{yxg_1(x,Q^2) + 2xg_2(x,Q^2)}{Y_+ 2xF_1(x,Q^2) - 2(1-y)F_L(x,Q^2)} \qquad (7)$$

Since $f_{A_\perp} \sim 0.013$ to $0.022$ for $E_p = 820\,\text{GeV}$ to $300\,\text{GeV}$ a measurement of $A_\perp$ at the same precision as for $A_\parallel$ would require luminosities much larger than the HERA design value. Due to this a measurement of $g_2(x,Q^2)$ in the HERA collider mode is not possible based on inclusive polarization asymmetries.

## 3  Parton Parametrizations

The dominant contribution to the structure function $g_1(x,Q^2)$ is of twist 2. It may be described in the parton model by

$$g_1(x,Q^2) = \frac{1}{2} \left\{ e_u^2 \sum_i \left[ \Delta u_i(x,Q^2) + \Delta \overline{u}_i(x,Q^2) \right] + e_d^2 \sum_i \left[ \Delta d_i(x,Q^2) + \Delta \overline{d}_i(x,Q^2) \right] \right\} \qquad (8)$$

with $e_u^2 = 4/9, e_d^2 = 1/9$ and $\Delta f(x,Q^2) = f^\uparrow(x,Q^2) - f^\downarrow(x,Q^2)$, where $f^{\uparrow\downarrow}$ are the parton densities at a given nucleon polarization. Here we assumed contributions from the light flavours only.

So far most of the constraints on the different flavour contributions to (8) are due to measurements of $g_1^p(x,Q^2)$ and $g_1^d(x,Q^2)$ at low values of $Q^2$ in the range of $0.01 \stackrel{<}{\sim} x$. The accuracy of the present data still leaves considerable freedom on the parametrization of $g_1^p(x,Q^2)$ and $g_1^n(x,Q^2)$ both in the small $x$ range and at larger values of $Q^2$ which are both accessible in possible later HERA experiments.

In figure 1 and 2 we compare four recent leading order parametrizations [5–8][1] of $g_1^p(x,Q^2)$ and $g_1^n(x,Q^2)$ in the range $10^{-4} < x < 1$ and $10 < Q^2 < 10^4\,\text{GeV}^2$. Although the $x$ shape of $g_1^{p,n}$ in the range $0.01 < x$ is quite similar for all parametrizations both the extrapolation to the small $x$ range and to higher $Q^2$ turn out to be rather different. This is due to the particular parametrization of the individual flavour contributions $x\Delta f_i$, and partly due to the starting point of the evolution choosen.

---

[1] If the LO $Q^2$ dependence was not provided by the authors of the respective parametrization it was derived from their Ansatz at $Q_0^2$ by LO QCD evolution using the CTEQ program [9]. I am indebted to Glenn Ladinsky for providing me with according parametrizations prior to publication [10]. In the case of parametrization [8] the set with $\alpha_G = 1$ was used choosing $\Delta u_s = \Delta \overline{u} = \Delta d_s = \Delta \overline{d} = 4\Delta s = 4\Delta \overline{s}$. The numerical illustrations below correspond to the 'standard scenario' in the case of parametrization [5] and the set 'gluon A' in ref. [6]. Note that in the case of ref. [6] a contribution $\propto \Delta G$ was added to eq. (8). In all other cases eq. (8) was used.

Whereas $g_1^p(x, Q^2)$ takes negative values for $x < 10^{-3}$ and shows a falling behaviour in the case of parametrization [5] for $10 < Q^2 < 10^4\,\text{GeV}^2$, it remains positive and rising for parametrization [7] in the same $Q^2$ range. For parametrization [6] the falling behaviour is only observed for $Q^2 > 10^3\,\text{GeV}^2$. Also parametrization [8] yields negative values in the range $x \sim \mathcal{O}(10^{-4}\ldots 10^{-3})$ at larger $Q^2$ which are, however, larger than those obtained from parametrization [5]. Moreover at still smaller $x$ values $g_1^p$ takes positive values again in the case of parametrization [8], while a falling behavior is obtained by [5, 6].

A similar uncertainty extrapolating from the kinematical range of the present measurements to the domain accessible at HERA is found comparing different parametrizations of $g_1^n$ (cf. figure 2.). However, the range in which $g_1^n$ takes positive values at large $x$ is about similar. The parametrization [8] predicts a rising behaviour of $g_1^n$ in the small $x$ range again.

A measurement of the $x$ shape of $g_1^p(x)$ at HERA should allow to distinguish between these parametrizations. Also the study of polarized $ed$ scattering would be interesting to constrain the behaviour of $g_1^n$ at smaller values of $x$ and larger values of $Q^2$.

## 4  Kinematical Range

Constraints on the kinematical range for neutral current deep inelastic scattering at HERA have been discussed in [11]. The different boundaries are implied by angle and energy cuts, bounds due to resolution effects [12], and large QED radiative corrections [13]. In figure 3 the accessible $x, Q^2$ range is shown for the HERA beam energies $E_e = 27.6\,\text{GeV}$ and $E_p = 820\,\text{GeV}$ demanding $E_{Jet} > 5\,\text{GeV}$, $\theta_{Jet} > 5^o$, $\theta_e < 175^o$, $0.01 < y < 0.9$, and $x < 0.7$. The values of $\langle Q^2 \rangle$ are correlated with $x$.[2] They are indicated by stars in figure 3 for the case of neutral current deep inelastic scattering. In the following we will investigate the sensitivity to measure the structure function $g_1(x, Q^2)$ in this range.

## 5  Accuracy of a measurement of $g_1^p(x, Q^2)$

In figure 4 the statistical accuracy of a measurement of $A_\parallel$ is illustrated as a function of $x$ assuming $\mathcal{L}_{int} = 30\,\text{pb}^{-1}$/beam polarization and $\lambda_e = \lambda_p = 0.8$. We used parametrization [6] as a reference value for the polarized parton densities, and [9] for the unpolarized densities[3].

The statistical error of $g_1(x, Q^2)$ measured from $A_\parallel$ is given by

$$\delta x g_1(x, Q^2) = \frac{1}{2\lambda_e \lambda_p} \frac{Y_+}{Y_-} \left[ 2xF_1(x, Q^2) - \frac{2(1-y)}{Y_+} F_L(x, Q^2) \right]$$
$$\times \left[ 2\mathcal{L}_{int} d^2\sigma_0 / dx dQ^2 \Delta x \Delta Q^2 \right]^{-1/2} \sqrt{1 - A_\parallel^2(x, Q^2)} \qquad (9)$$

where $d^2\sigma_0/dxdQ^2$ denotes the unpolarized differential scattering cross section, $\Delta x \Delta Q^2$ is the bin size, and $\mathcal{L}_{int}$ the integrated luminosity per beam polarization. $\delta g_1^p(x, Q^2)$ does only weakly depend on the value of $g_1(x, Q^2)$ itself as long as $A_\parallel^2 \ll 1$.

---

[2]For $Q^2 \gtrsim 1000\,\text{GeV}^2$ the neutral current deep inelastic scattering cross section contains also contributions due to $\gamma Z$ interference and $Z$ excange terms which are related to new structure functions. As in the case of the measurement of $F_2(x, Q^2)$ (cf. [2]) these terms may be delt with as corrections in the measurement of $g_1^p(x, Q^2)$.

[3]In the $x$ and $Q^2$ range considered the different parametrizations for unpolarized parton densities agree to a wide extent. The contribution due to $F_L$ was disregarded in the numerical calculation of $A_\parallel$.

It turns out that at the value of $\mathcal{L}_{int}$ considered the product of the beam polarizations should take values of
$$\lambda_e \lambda_p \gtrsim 0.5 \qquad (10)$$
to obtain a sufficient resolution for $g_1^p$.

In figure 5 the accuracy of a $x$ shape measurement of $g_1^p(x, Q^2)$ in the kinematical range of HERA (cf. figure 3) is shown. The values of $g_1(x)$ represent averages over $Q^2$. With rising values of $x$ $\langle Q^2 \rangle$ rises. Assuming the parametrization ref. [6] as a reference value relative errors for $g_1^p(x)$ between 20% and 80% in the $x$ range from 0.5 to $5 \cdot 10^{-4}$ at $3500 > Q^2 > 15\,\mathrm{GeV}^2$ are obtained under the above conditions. Predictions for $g_1^p(x, \langle Q^2 \rangle)$ by other parametrizations are shown for comparison. We compare possible measurements at proton beam energies $E_p = 820\,\mathrm{GeV}$ and $E_p = 300\,\mathrm{GeV}$. The statistical precision of the measurement improves with rising values of $x$ and is larger in the case of the low energy option at which the range down to $x \sim 10^{-3}$ can be probed still. The measurement of $g_1^p$ in this kinematical range will allow to further constrain existing parametrizations of polarized parton densities. Particularly it will be interesting to see whether $g_1^p(x, Q^2)$ takes negative values in the small $x$ range.

In figure 6 the statistical accuracy of $g_1^p(x, \langle Q^2 \rangle)$ is illustrated in the range $x > 0.01$. Differences between the parametrizations [5–8] are still observed. The size of the scaling violations in LO in this $x$ range is shown comparing the values of $g_1^p(x)$ for $Q^2 = \langle Q^2 \rangle$ and $Q^2 = 4\,\mathrm{GeV}^2$, the range of the data from fixed target experiments. For $x < 0.1$ the scaling violations of $g_1^p(x, Q^2)$ are of the size of $\delta g_1^p(x)$ for a measurement under the conditions mentioned above. Towards larger $x$ values the scaling violations shrink due to the fix point at $x \sim 0.15$. Thus, to see scaling violations of $g_1^p(x, Q^2)$ in a clear way requires a higher luminosity.

The statistical accuracy improves as $\langle Q^2 \rangle$ becomes smaller since the differential scattering cross sections behave $\sim 1/Q^4$. In figure 7 we illustrate the statistical precision of a possible measurement at $E_p = 30\,\mathrm{GeV}$ and $E_e = 3\,\mathrm{GeV}$ [14] whith $\lambda_e = \lambda_p = 0.8$, $\mathcal{L}_{int} = 100\,\mathrm{pb}^{-1}$/polarization and applying the kinematical cuts described in section 4. The range of $Q^2$ at these beam energies is rather small and extends from 4 to $12\,\mathrm{GeV}^2$ only. A very precise shape measurement of $g_1^p(x, \langle Q^2 \rangle)$ is possible in the range $x > 10^{-2}$ at this option.

# 6 Conclusions

The measurement of the structure function $g_1^p(x, Q^2)$ in the HERA collider mode with both longitudinally polarized electron and proton beams would allow to probe the behaviour of this structure function at smaller values of $x$ and larger values of $Q^2$ compared to measurements possible in fixed target experiments. $g_1^p(x)$ can be measured at a statistical precision of 80% to 20% in the range of $5 \cdot 10^{-4} < x < 0.5$ correlated with values of $15 < \langle Q^2 \rangle < 3500\,\mathrm{GeV}^2$ for an integrated luminosity of $\mathcal{L}_{int} = 30\mathrm{pb}^{-1}$/beam polarization and polarization values of $\lambda_e = \lambda_p = 0.8$. The measurement of $g_1^p$ using a lower proton energy, $E_p = 300\,\mathrm{GeV}$, leads to an improvement of the statistical accuracy without a significant restriction in the $x$ range. Further constraints on the $x$ behaviour of $g_1^p$ can be obtained by these measurements. A detailed investigation of scaling violations of $g_1(x, Q^2)$ requires a larger integrated luminosity.

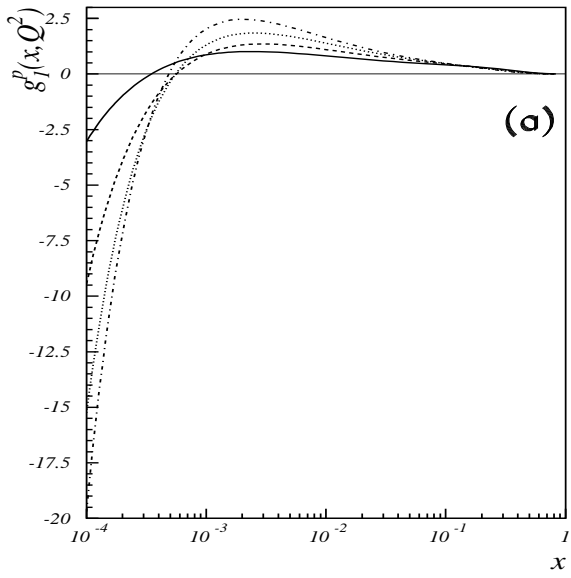
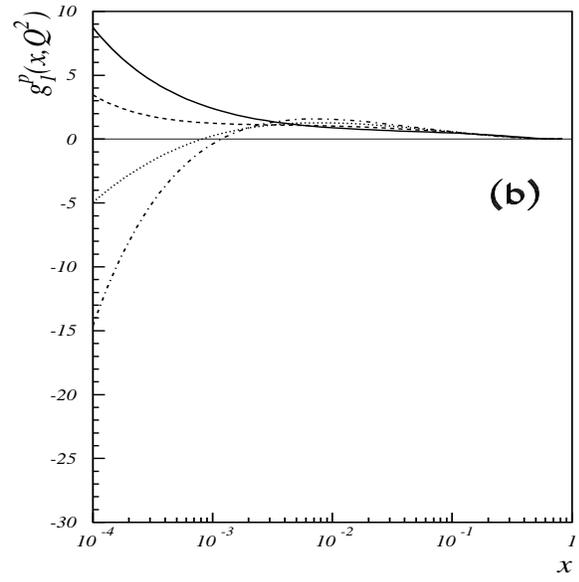
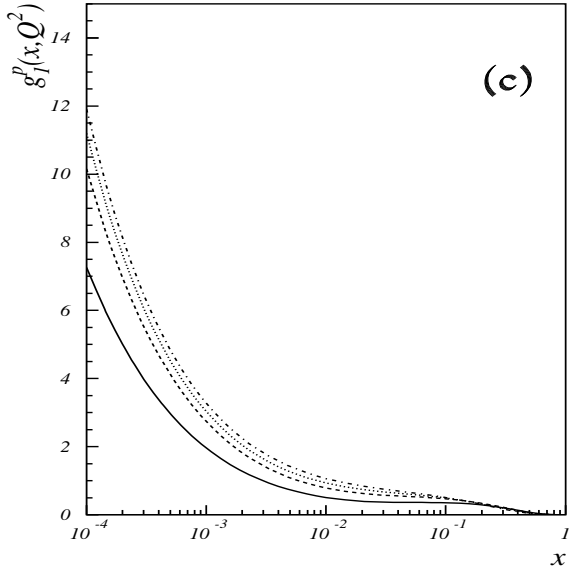
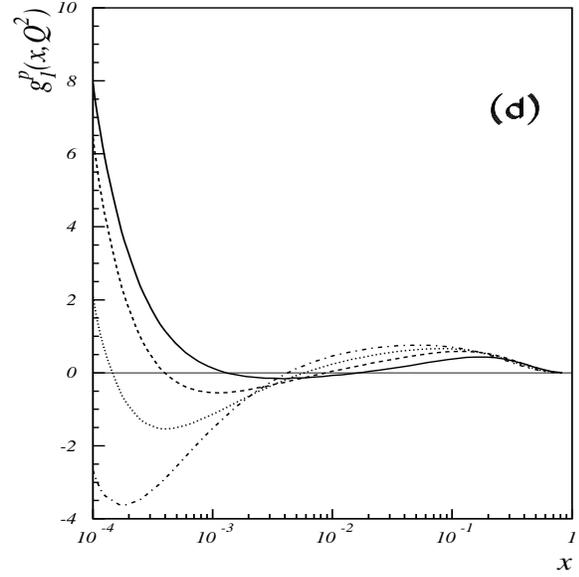

Figure 1: The structure function $g_1^p(x, Q^2)$ in the range $x > 10^{-4}$. Full line: $Q^2 = 10\,\text{GeV}^2$, dashed line: $Q^2 = 10^2\,\text{GeV}^2$, dotted line: $Q^2 = 10^3\,\text{GeV}^2$, dash–dotted line: $Q^2 = 10^4\,\text{GeV}^2$. The parametrizations are: (a) ref. [5], (b) ref. [6], (c) ref. [7], (d) ref. [8].

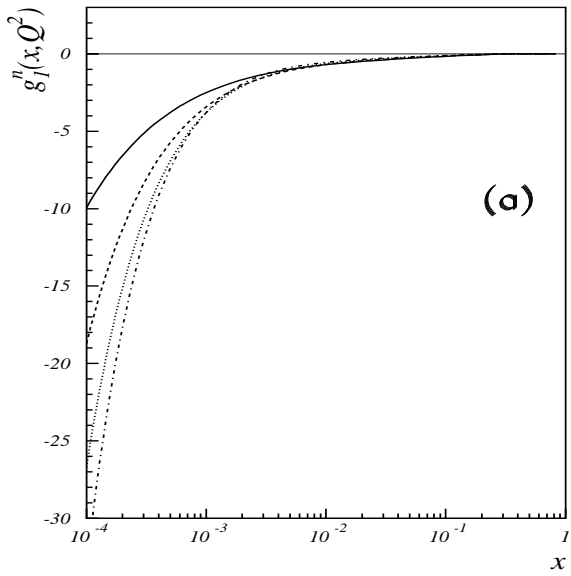
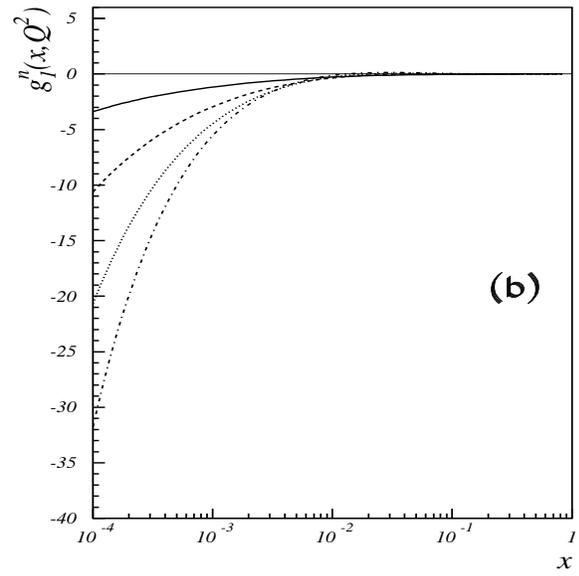
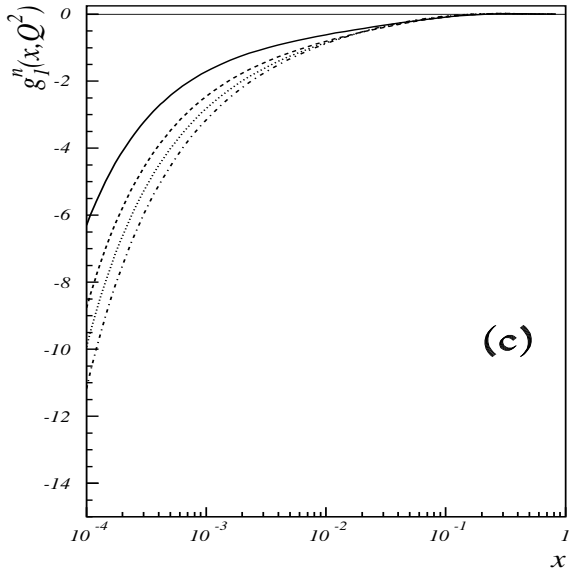
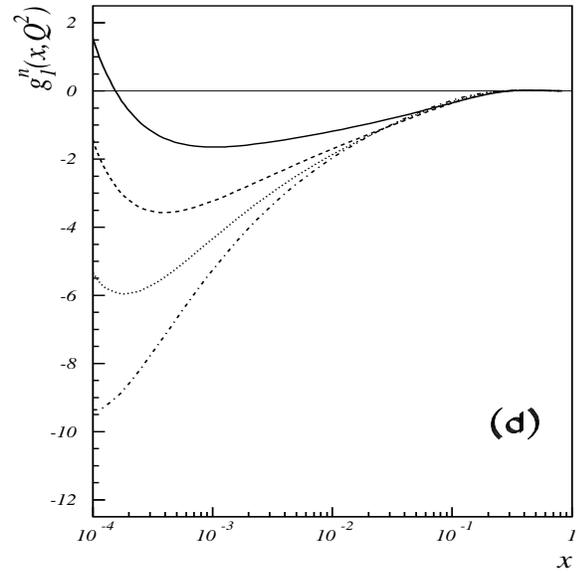

Figure 2: The structure function $g_1^n(x, Q^2)$ in the range $x > 10^{-4}$. Full line: $Q^2 = 10\,\text{GeV}^2$, dashed line: $Q^2 = 10^2\,\text{GeV}^2$, dotted line: $Q^2 = 10^3\,\text{GeV}^2$, dash–dotted line: $Q^2 = 10^4\,\text{GeV}^2$. The parametrizations are: (a) ref. [5], (b) ref. [6], (c) ref. [7], (d) ref. [8].

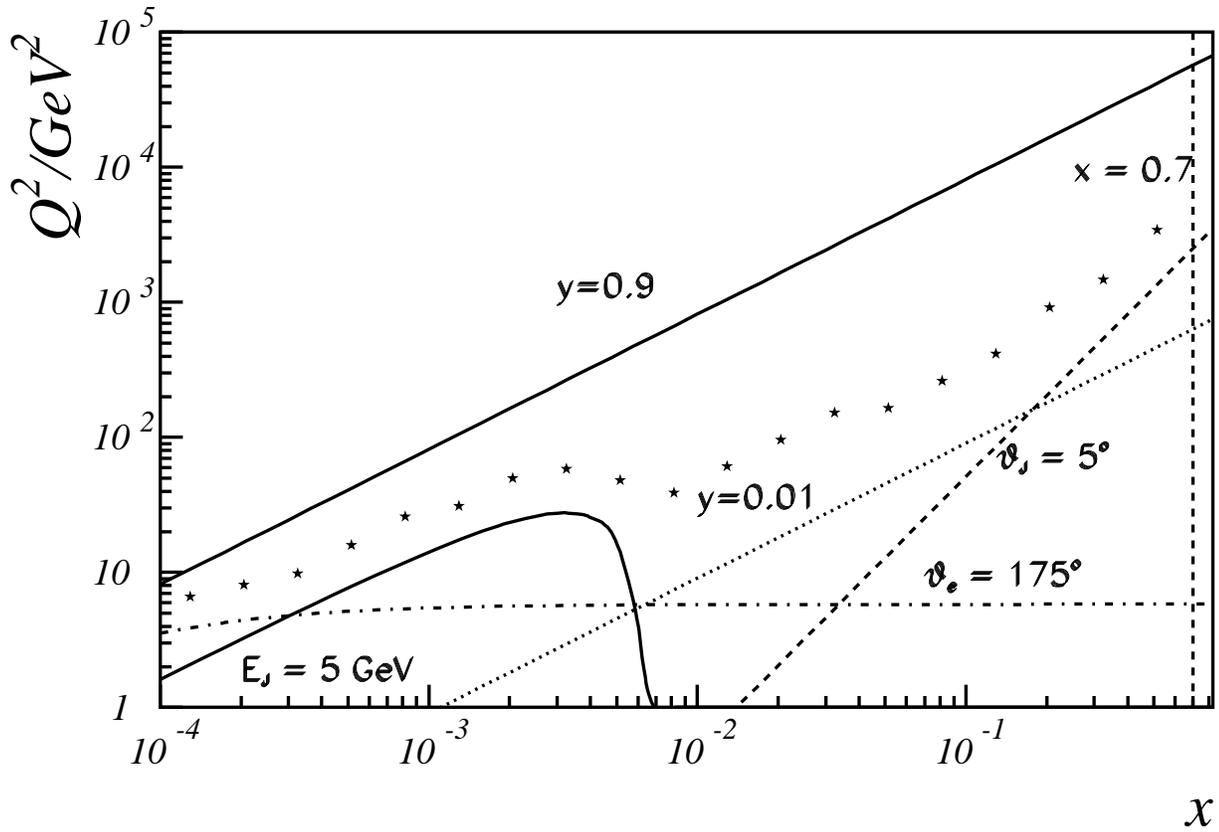

**Figure 3**: The accessible kinematical range for neutral current deep inelastic scattering at HERA; $E_p = 820\,\text{GeV}$, $E_e = 27.6\,\text{GeV}$. The stars indicate the values of $\langle Q^2 \rangle$ at a given value of $x$ for neutral current deep inelastic scattering.

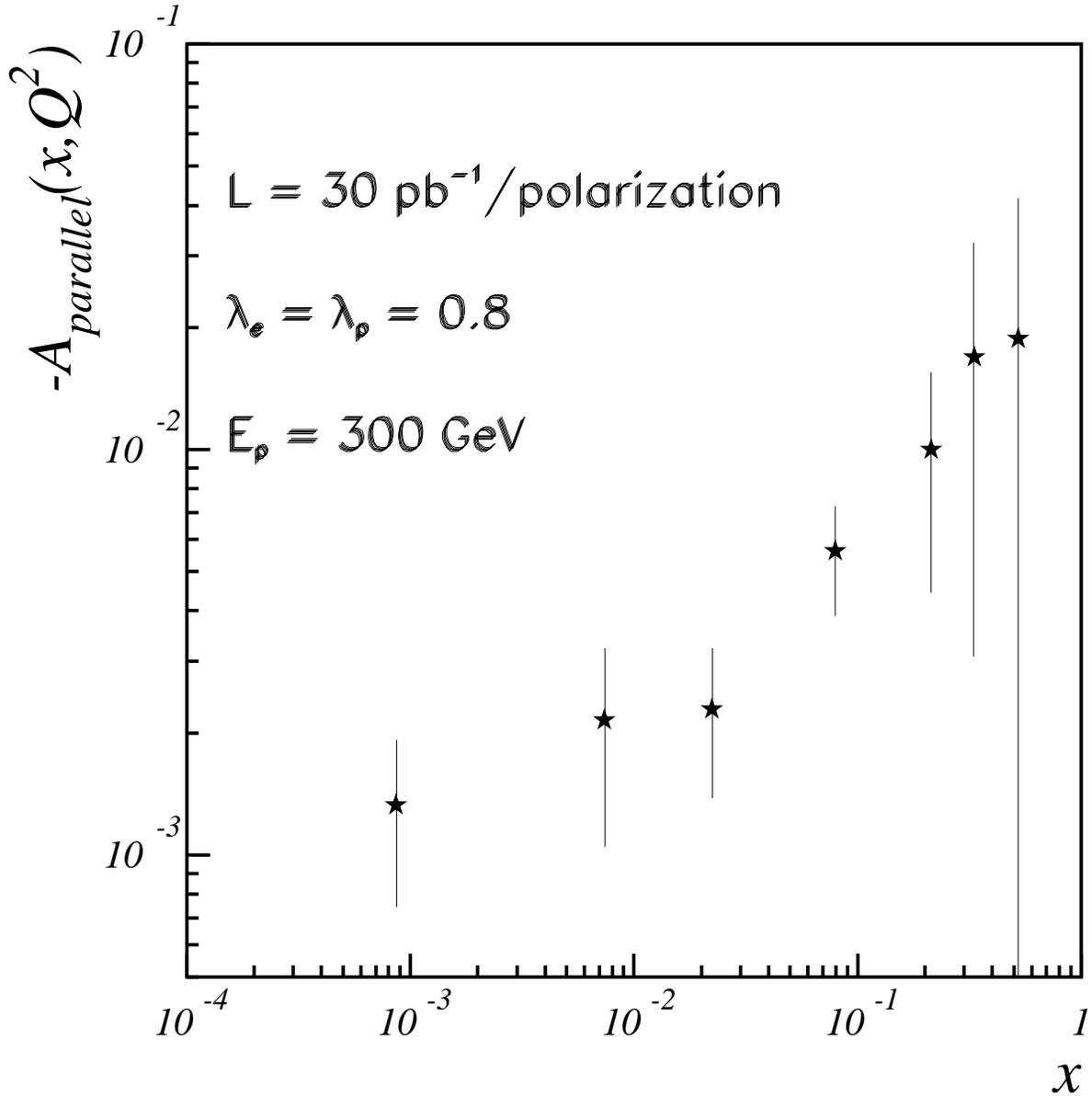

**Figure 4**: Statistical precision of a measurement of $-A_{\|}(x, \langle Q^2 \rangle)$ in the kinematical domain of HERA for $E_p = 300$ GeV. The data points represent averages over the accessible $Q^2$ range and were calculated using the parametrizations [6, 9].

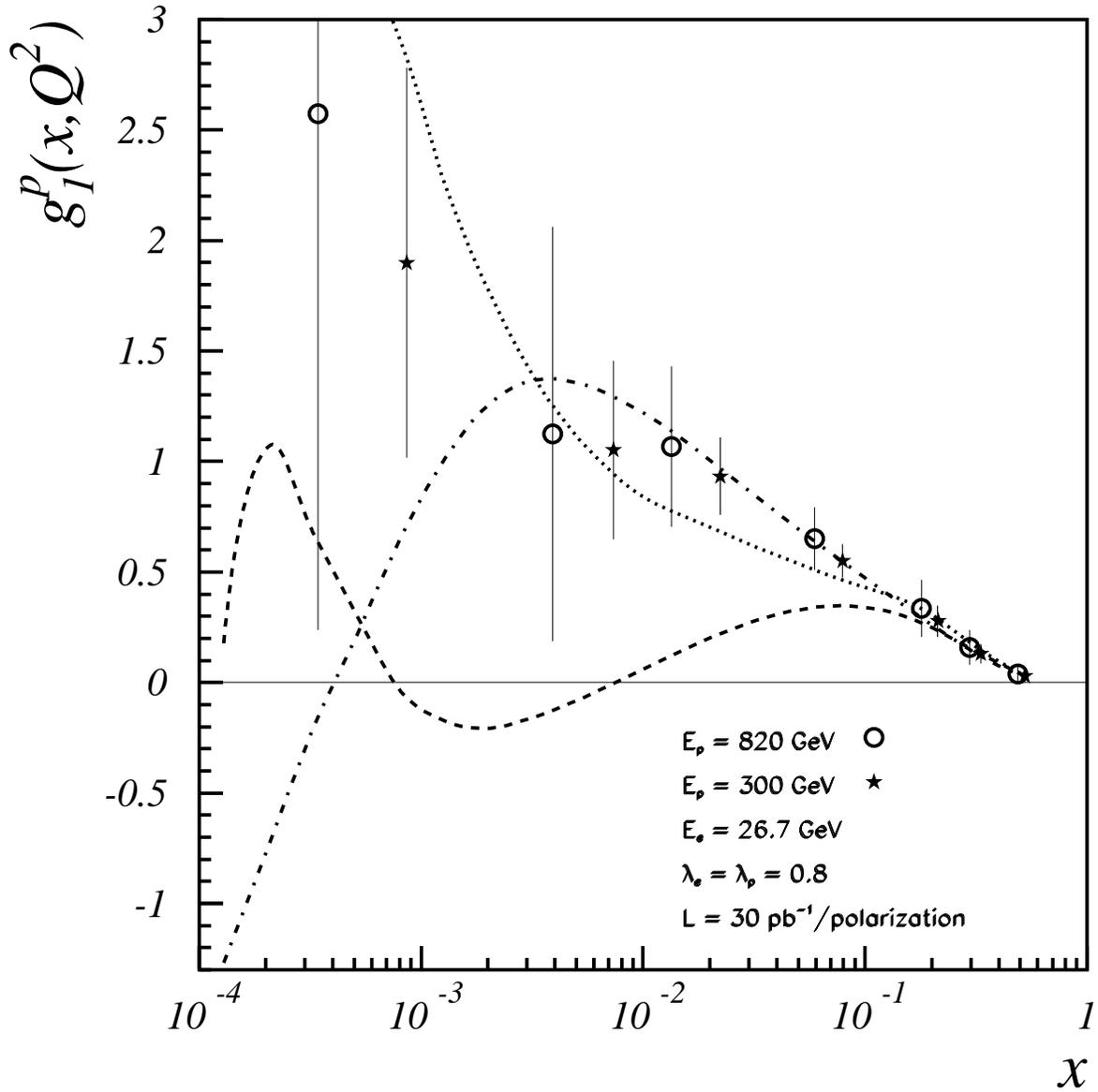

Figure 5: Statistical precision of a measurement of $g_1^p(x, Q^2)$ in the kinematical domain of HERA. The data points represent averages over the accessible $Q^2$ range and were calculated using the parametrization [6]. The dashed, dotted line, and dash–dotted line correspond to the values of $g_1^p(x, \langle Q^2 \rangle)$ for the parametrizations [8], [7], and [5], respectively.

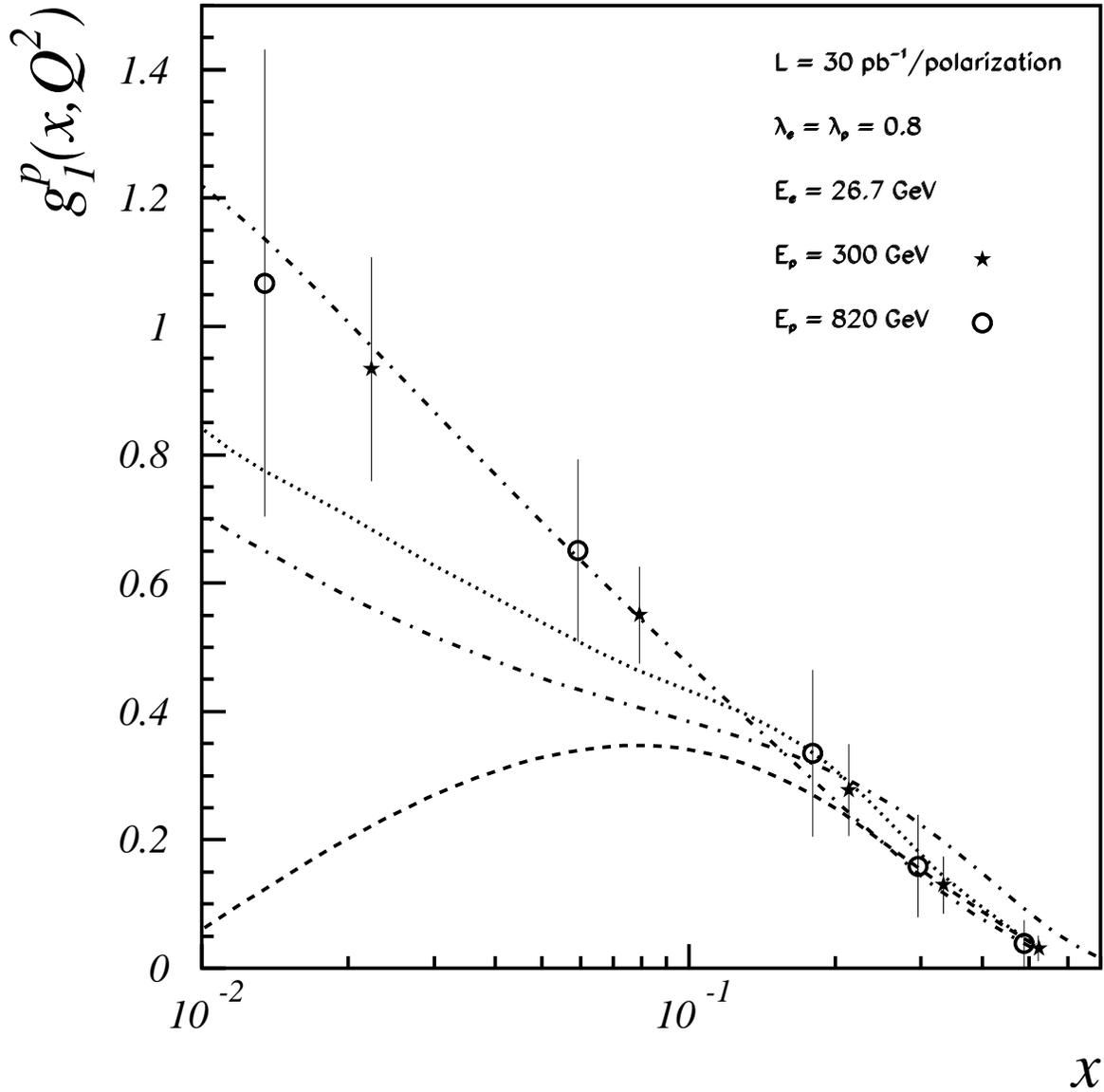

Figure 6: Statistical precision of a measurement of $g_1^p(x, Q^2)$ in the kinematical domain of HERA at larger values of $x$. The data points represent averages over the accessible $Q^2$ range and were calculated using the parametrization [6]. The dashed, dotted, and upper dash–dotted line correspond to the values of $g_1^p(x, \langle Q^2 \rangle)$ for the parametrizations [8], [7], and [5], respectively. The lower dash–dotted line shows $g_1^p(x, Q_0^2)$ for $Q_0^2 = 4\,\text{GeV}^2$ for parametrization [5].

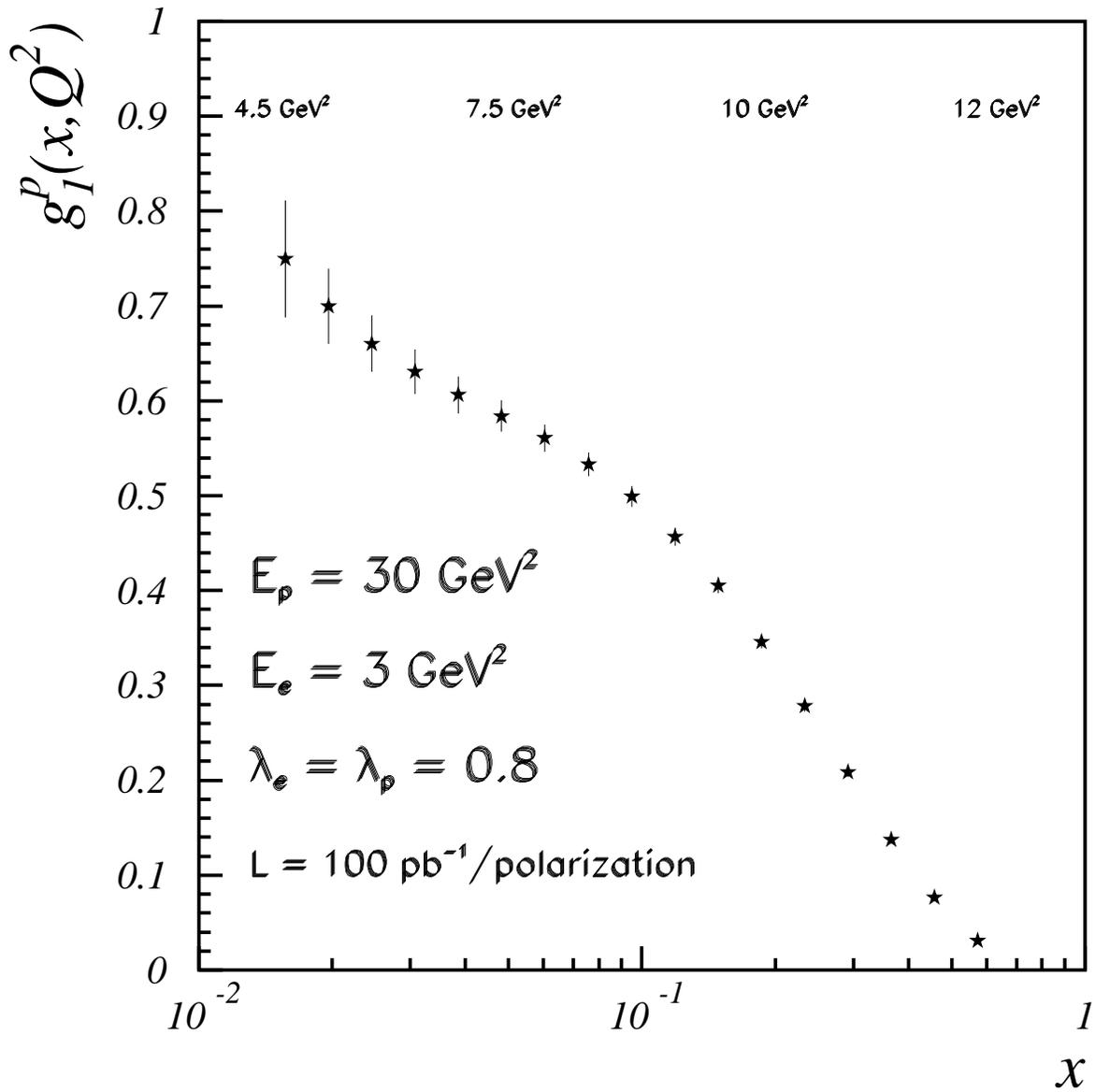

Figure 7: Statistical precision of a measurement of $g_1^p(x, \langle Q^2 \rangle)$ for $ep$ scattering at $E_e = 3\,\text{GeV}$ and $E_p = 30\,\text{GeV}$ assuming the same kinematical cuts as at HERA.